\documentclass[prl,aps,twocolumn,showpacs,floatfix,longbibliography]{revtex4-2}
\usepackage{amsfonts}
\usepackage{amssymb}
\usepackage{hyperref}
\usepackage{graphicx}
\usepackage{dcolumn}
\usepackage{bm,amsmath,verbatim}
\usepackage{mathrsfs}
\usepackage{color}
\usepackage[T1]{fontenc}
\usepackage[latin9]{inputenc}
\usepackage{booktabs}

\hypersetup{colorlinks,
	linkcolor=blue,          citecolor=blue,        filecolor=blue,      urlcolor=blue           }

\begin{document}
			
\title{Transitions from Composite Fermi Liquid to Moore-Read States in Weyl Semimetals}
\author{Jiong-Hao Wang$^{1}$}
\author{Yong Xu$^{1,2}$}
\email{yongxuphy@tsinghua.edu.cn}
\affiliation{$^{1}$Center for Quantum Information, IIIS, Tsinghua University, Beijing 100084, People's Republic of China}
\affiliation{$^{2}$Hefei National Laboratory, Hefei 230088, People's Republic of China}	

	\begin{abstract}
Weyl semimetals represent a significant class of topological gapless materials in three dimensions and 
have been shown to exhibit three-dimensional quantum Hall effect. However, existing research 
mainly focuses on scenarios without interactions. Recent studies suggest that the fractional quantum Hall 
effect can arise in a Weyl semimetal with a one-third filled Landau level under a magnetic field.
However, it remains unclear whether more exotic states, such as composite Fermi liquid and 
Moore-Read states, can appear at half filling.
Here we surprisingly find the existence of composite Fermi liquid, Moore-Read states and charge density waves 
in the same Weyl-orbit-based Landau level of a Weyl semimetal and their transitions induced by varying
the distance between Weyl points.
We attribute these transitions to a significant change in the single-particle 
wave functions of the Landau level as the distance between Weyl points is varied.
Finally, we demonstrate that a transition from composite Fermi liquid
to Moore-Read states can be induced by tuning the direction of a magnetic field, a process that is more 
experimentally accessible.
	\end{abstract}
	\maketitle

\section{Introduction}

Weyl semimetals~\cite{jia2016NM,burkov2016NM,armitage2018RMP,xu2019FP,lv2021RMP}, 
 a prominent class of three-dimensional (3D) gapless topological materials,
have attracted intense research interest in recent years~\cite{wan2011PRB,yang2011PRB,burkov2011PRL,xu2011PRL,witczak2014PRL,jian2015PRB,potter2014NC,bergholtz2015PRL,weng2015PRX,
	yang2015PRL,chen2015PRL,xu2015PRL,soluyanov2015Nature,morimoto2016SciRep,pixley2016PRX,ishizuka2016PRL,carlstrom2018PRB,yang2019PRL,
wang2020PRL,yao2023PRX,nandy2024PRB}.
Their band structures feature pairs of Weyl points that possess opposite topological charges and
exhibit linear dispersion nearby. 
According to the bulk-boundary correspondence, the topology of the band structures leads to  
open Fermi surfaces (known as Fermi arcs) on
the surface of a sample that connects projections of Weyl points with opposite charges.
Due to the topological property, 
Weyl semimetals exhibit topological anomalous Hall effect~\cite{yang2011PRB,burkov2011PRL}.
In addition, it has been shown that 
3D quantum Hall effect can arise in Weyl semimetals or Dirac semimetals under a magnetic 
field~\cite{wang2017PRL,zhang2017NC,uchida2017NC,schumann2018PRL,zhang2019Nature,
lu2019NSR,li2020PRL,nishihaya2021NC,chen2021PRL,zhang2021NRP,li2021npjQ,chang2022CPB,xiong2022PRB,zhangxr2023PRB,wang2023arXiv,nakazawa2024JPSJ,qin2024PRB,kong2024PRB,
chang2024PRB,zhang2024SciBull,ye2024PRL,wang2025PRB} 
via Weyl orbits~\cite{potter2014NC}, a phenomenon that has recently been confirmed experimentally~\cite{zhang2019Nature}.
However, these studies mainly focus on scenarios without interactions.

It is well known that the interplay between topology and interactions can lead to exotic topological phases of matter 
in two dimensions (2D), such as fractional quantum Hall (FQH) states~\cite{jain1992AdvPhys,stormer1999RMP,
	stormer1999nobel,murthy2003RMP,hansson2017RMP}.
A typical example is the Laughlin FQH state at $1/m$ filling ($m$ is an odd integer) 
that exhibits properties such as 
fractionally quantized Hall conductance as well as anyonic excitations with fractional charges 
and fractional statistics. 
Additionally, at various other filling factors with odd denominators, 
hierarchy FQH effect~\cite{hansson2017RMP} can arise.
They can be understood heuristically as an integer quantum Hall 
effect of composite fermions composed of bare electrons bound to an even number of vortices~\cite{jain2007}. 
Beyond the odd denominator case, fascinating physics emerges at filling factors with even denominators. 
For example, at half filling of the lowest Landau level, there appears 
composite Fermi liquid (CFL)~\cite{halperin1993PRB} where composite fermions feel 
zero effective magnetic fields.
For the first excited Landau level, the Moore-Read (MR) state~\cite{read1991NPB,read1999PRB} emerges,
which belongs to the non-Abelian FQH effect featuring 
non-Abelian anyons~\cite{read1991NPB,read1999PRB,nayak2008RMP}
with potential applications in topological quantum computation~\cite{nayak2008RMP}.
These phenomena have been extensively explored in 2D systems.
Interestingly, recent developments have shown that Weyl semimetals can also exhibit 
FQH effect at one-third filling~\cite{wang2025PRB}, 
based on the fact that Weyl semimetals can support Landau levels arising from Weyl orbits under a magnetic field. 
However, it remains an open question whether CFL and MR states can
also appear in Weyl semimetals at half filling. 
  
Here we theoretically study the ground state properties of electrons in Weyl semimetals under 
a magnetic field in the presence of Coulomb interactions, when a Landau level is half filled.
We surprisingly observe a transition from the CFL to MR states and then to a charge density wave (CDW) in the same Landau level
as we increase the separation between Weyl points.
The transition is attributed to a significant change in the single-particle 
wave functions of the Landau level as the separation between Weyl points is varied.
We identify the CFL phase by analyzing its low energy states in the many-body energy spectrum, which resemble those of the 2D CFL,
and by examining the guiding center structure factor that reveals the Fermi surface of composite fermions.
The MR states are identified through momentum-resolved low energy states in the energy spectrum and 
by counting eigenvalues below the gap in particle entanglement spectra, both of which
are consistent with the (2,4) counting rule for MR states.
As the separation between the Weyl points further increases, we observe a high degeneracy in the energy spectrum 
and prominent isolated peaks in the structure factor, indicating the presence of the CDW phase.
Finally, we demonstrate that tuning the direction of a magnetic field can induce a transition from the CFL
to MR states, a process that is more experimentally accessible.

\section{Model}	
\begin{figure}[t]
 	\includegraphics[width=3.4in]{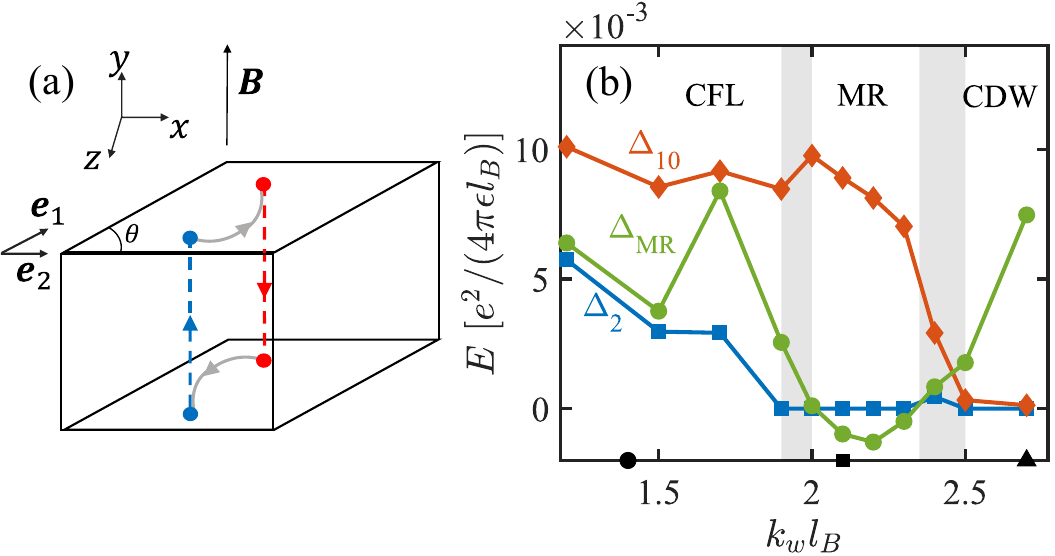}
 	\caption{(a) Schematic illustration of Fermi arcs (gray curves) and Weyl
 		orbits developed under a magnetic field $\bm{B}$ for the Hamiltonian~(\ref{Ham1}) 
 		(see Appendix A). 
 		We consider a geometry characterized by a parallelogram in the horizontal plane, defined by two unit vectors 
 		$\bm{e}_1$ and $\bm{e}_2$ with an angle $\theta$.
 		(b)
 		The phase diagram for the interacting Hamiltonian~(\ref{Ham2}) at half filling with respect to half the 
 		distance between two Weyl points ($k_w$). Three phases including the CFL, the MR state and the CDW
 		are identified. The energy differences $\Delta_2$ (blue line), $\Delta_{10}$ ( 
 		red line) and $\Delta_{\textrm{MR}}$ (green line) are defined in the main text.
 		The critical regions are indicated by the gray color.  
 		The black circle, square and triangle on the $k_w$ axis denotes 
 		the points with $k_w l_B=1.4,2.1,2.7$, respectively. 
		Here, $l_B=\sqrt{\hbar/(eB)}$ is the magnetic length. 
		The Landau levels are calculated for the parameters
 		$\widetilde{M}=M/l_B^2=0.03E_0,\widetilde{D}_1=D_1/l_B^2=0.025E_0,
 		\widetilde{D}_2=D_2/l_B^2=0.005E_0$ and $\widetilde{A}=A/l_B=0.04E_0$,
 		with $E_0$ being the unit of energy.
 		This set of parameters is used throughout the paper unless stated otherwise.
 	}
 	\label{Fig1}
 \end{figure}

We start by considering a single-particle Weyl semimetal model 
described by the following continuous Hamiltonian~\cite{wang2017PRL,wang2025PRB}
\begin{eqnarray}\label{Ham1}
	H(\bm{k})&=&M[k_w^2-(k_x^2+k_y^2+k_z^2)]\sigma_z+A(k_x \sigma_x +k_y\sigma_y) \nonumber \\
	&&+[D_1k_y^2+D_2(k_x^2+k_z^2)]\sigma_0,
\end{eqnarray} 	
where $\sigma_\mu$ with $\mu=0,x,y,z$ denote identity and Pauli matrices, respectively, 
$\bm{k}=(k_x,k_y,k_z)$ is the momentum, and $M$, $A$, $D_1$, $D_2$, and $k_w$ 
are system parameters.
The model supports a pair of Weyl points at $(0,0,\pm k_w)$. 
Under a magnetic field $\bm{B}=(0,B,0)$ along $y$, 
the Weyl orbit formed by 
Fermi arcs [see Fig.~\ref{Fig1}(a)] gives rise to Landau levels.
We here focus on the Landau level so that the total Chern number of all bands below this Landau level 
sums to zero
(see Appendix A),
and consider the case that this Landau level is half filled.
We now investigate the possible phases in the half-filled Landau level with the Coulomb interaction 
between electrons
in a sample geometry exhibiting a parallelogram shape with a $\theta=\pi/3$ inner angle 
in the horizontal plane [see Fig~\ref{Fig1}(a)].
The interaction under periodic boundary conditions (PBCs) along $\bm{e}_1$ and $\bm{e}_2$ 
and open boundary conditions (OBCs) along $y$ reads 
\begin{equation}\label{Coulomb}
	V(\bm{r})=\sum_{t,s=-\infty}^{+\infty}\frac{e^2}{4\pi\epsilon}\frac{1}
	{|\bm{r}+tL_1\bm{e}_1+sL_2\bm{e}_2 |},
\end{equation} 
where $e$ is the electron charge, $\epsilon$ is the dielectric constant, and 
$\bm{e}_1$ and $\bm{e}_2$ are unit vectors as shown in Fig.~\ref{Fig1}(a) with $L_1$ and $L_2$ being the 
system's length along these directions, respectively.
Following the conventional single-band projection approximation in the study of Landau level physics~\cite{yoshioka1983PRL,rezayi1999PRL,rezayi2000PRL,haldane2000PRL,wang2025PRB}, which is the only approximation in our numerical calculations,
we project the interactions onto the Landau level of interest 
and arrive at the interacting Hamiltonian
 \begin{equation}\label{Ham2}
	\hat{H}_I=\sum_{\bm{k}_1,\bm{k}_2,\bm{k}_3,\bm{k}_4}C_{\bm{k}_1 \bm{k}_2\bm{k}_3\bm{k}_4}
	\hat{c}^\dagger_{\bm{k}_1}\hat{c}^\dagger_{\bm{k}_2}
	\hat{c}_{\bm{k}_3}\hat{c}_{\bm{k}_4}.
\end{equation} 
Here $\hat{c}_{\bm{k}}$ ($\hat{c}^\dagger_{\bm{k}}$) is the annihilation (creation) operator of
a single-particle state at a two-dimensional (2D) momentum $\bm{k}=k_1\bm{g}_1+k_2\bm{g}_2$ 
in the considered Landau level, where 
$k_i=0,1,...,N_i-1$ for $i=1,2$ with $ {N_1 N_2} $ equal to 
the degeneracy of the Landau level,
and $N_j\bm{g}_j$ ($j=1,2$) play the roles of reciprocal primitive vectors 
such that $\bm{g}_i\cdot \bm{e}_j=\delta_{ij}2\pi/L_i $
(see Appendix B).
In numerical calculations, we drop the 
quadratic terms that contribute a constant value~\cite{wang2025PRB}, and focus solely  
on the pure interacting Hamiltonian (\ref{Ham2}).
For simplicity, we consider the geometry with $N_1/N_2=L_1/L_2$ so that the 
Brillouin zone is a hexagon (see Appendix B).
We consider $N_1=N_2=4$ for eight electrons, $N_1=4, N_2=5$ for ten electrons,
and $N_1=4,N_2=6$ for twelve electrons.

 \section{Phase diagram}
\begin{figure}[t]
  	\includegraphics[width=3.4in]{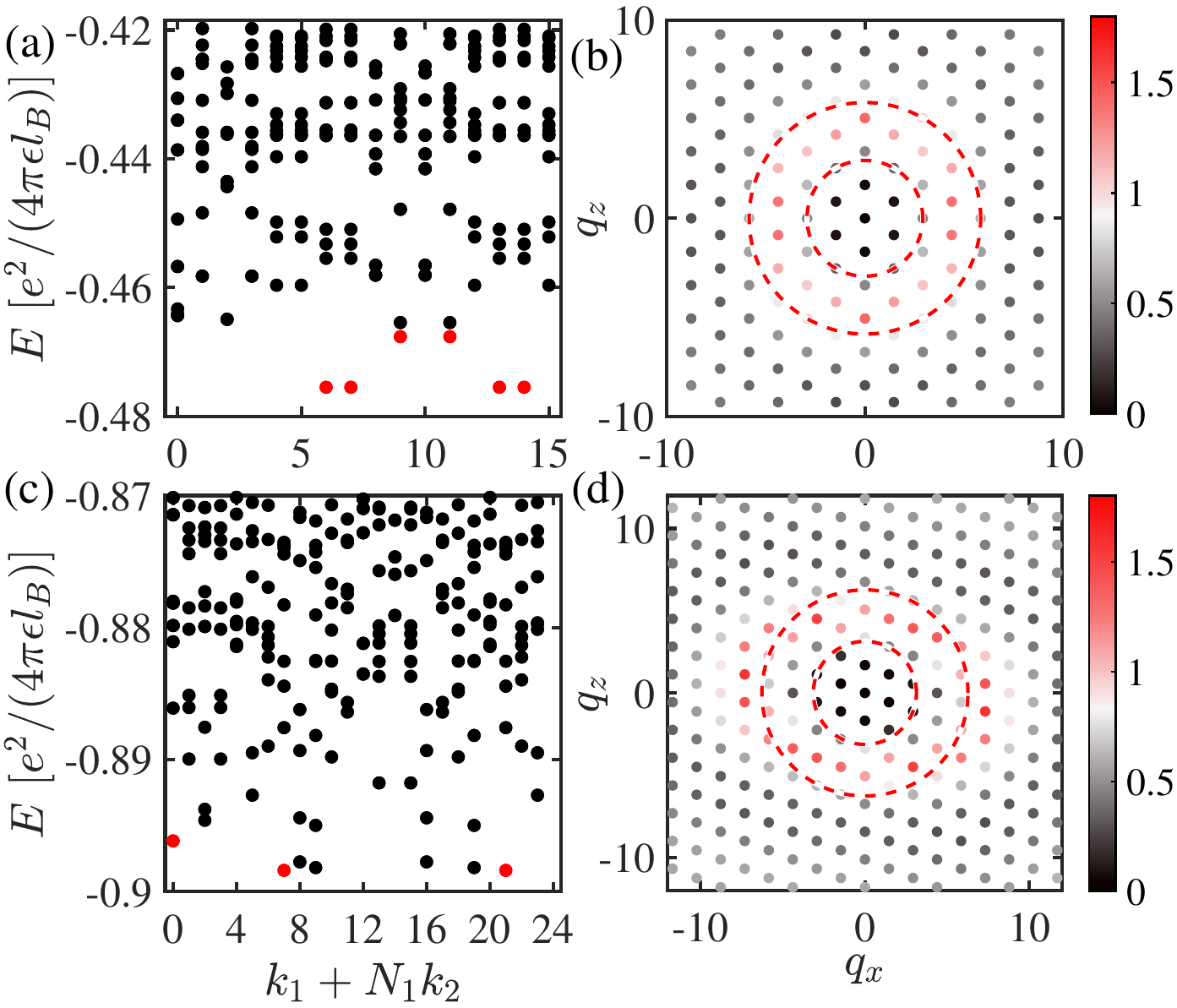}
  	\caption{The many-body energy spectrum for (a) eight and (c) twelve electrons at $k_w=1.4/l_B$. 
  		The low energy states similar to the 2D CFL are colored red with each red point representing
  		two degenerate states. 
  		The guiding center static structure factor $S(\bm{q})$ of the ground states for (b) eight and (d) twelve electrons.
  		The inner red circle describes the Fermi surface of free electrons with the radius $k_F$ and the outer one 
  		indicates $2k_F$.
  	}
  	\label{Fig2}
  \end{figure}

\begin{figure}[t]
  	\includegraphics[width=3.4in]{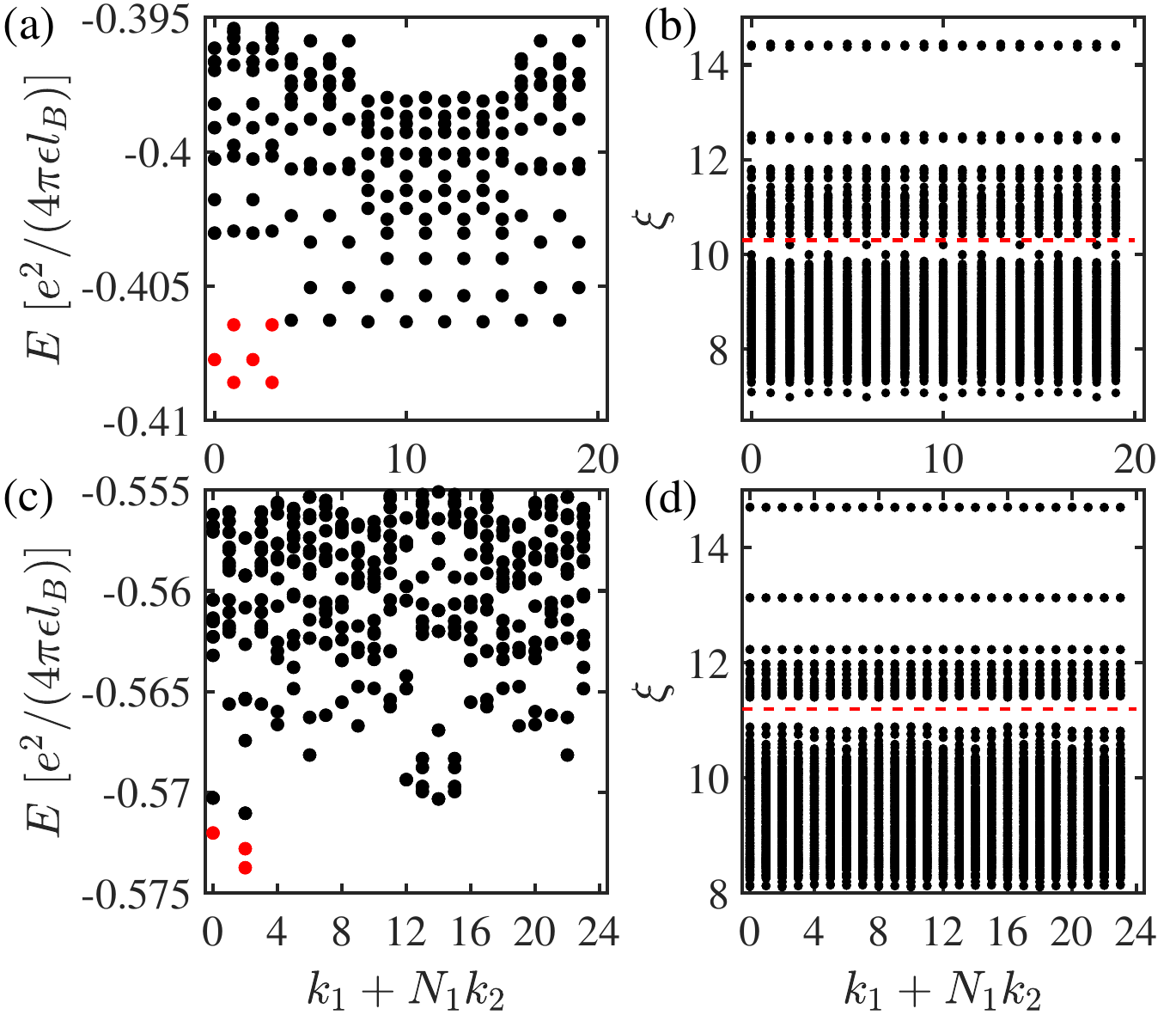}
  	\caption{The many-body energy spectrum for (a) ten and (c) twelve electrons when $k_w=2.1/l_B$. 
  		The MR states are indicated by the red color. 		
  		The particle entanglement spectrum for (b) ten and (d) twelve electrons.
  		The number of states below the red dashed line is consistent with the counting rule for the MR states.
  } 
  	\label{Fig3}
  \end{figure}

\begin{figure}[t]
	\includegraphics[width=3.4in]{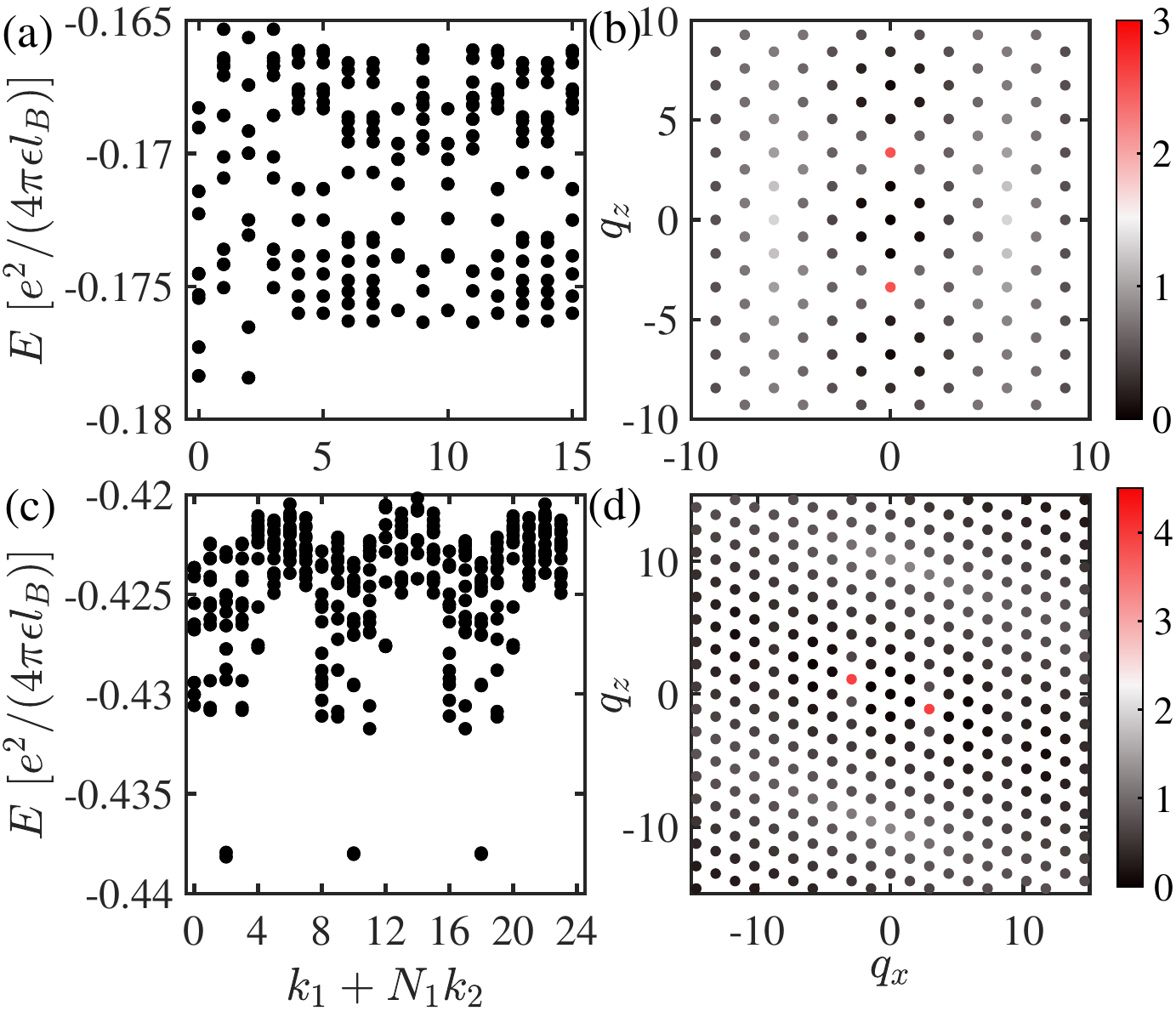}
	\caption{The many-body energy spectrum for (a) eight and (c) twelve electrons when $k_w=2.7/l_B$. 
		The guiding center static structure factor $S(\bm{q})$ of the ground states for (b) eight and (d) twelve electrons.
	}
	\label{Fig4}
\end{figure}

We find three strongly-correlated phases: CFL, MR states
and CDW, as shown in Fig.~\ref{Fig1}(b). We see that the CFL first transitions to 
the MR states and then to the CDW as the distance between the two Weyl points increases.
The three distinct phases are initially identified by calculating energy differences for a system with
twelve electrons. A detailed analysis of these phases will be presented in the following sections. 
Specifically, since one of the ground states in the MR phase and CDW phase lie in the momentum sector  
with $k_1+N_1k_2=2$ [see Fig.~\ref{Fig3}(c)] and $k_1+N_1 k_2=10$ [see Fig.~\ref{Fig4}(c)], respectively, 
the energy difference $\Delta_2$ ($\Delta_{10}$) between
the lowest energy in the former (latter) momentum sector and the ground state energy of the
system is able to reveal the transition from the CFL (MR state) to the MR state (CDW).
Indeed,
we see that as $k_w$ increases, $\Delta_2$ drops to zero near $k_w=1.9$, indicating the transition 
to the MR state.  
Note that the ground states for the CFL phase do not appear in these momentum sectors [see Fig.~\ref{Fig2}(c)].
The sharp decline of the red line near $k_w=2.4$ implies a transition to the CDW phase.
In addition, we plot $\Delta_{\textrm{MR}}$ defined as the highest energy of the MR manifold minus the lowest 
 energy of the other states ($-\Delta_{\textrm{MR}}$ indicates the energy gap in the MR phase).
Here the MR manifold is composed of two lowest energy states with  $k_1+N_1k_2=0$ and four lowest energy states with  $k_1+N_1k_2=2$,
which are just the MR states in the MR phase [see Fig.~\ref{Fig3}(c)].
When $\Delta_{\textrm{MR}}<0$, it signifies a finite gap above the MR manifold, corresponding to the MR region with
$\Delta_2=0$ and $\Delta_{10}>0$ in Fig.~\ref{Fig1}(b).

In Appendix C, we show in detail that the phase transition arises due to the dramatic change in 
the single-particle wave functions of the considered Landau level as $k_w$ increases. 
In the following, we will investigate the properties of the three phases in detail.   
 
  \section{Composite Fermi liquid}
We numerically diagonalize the Hamiltonian (\ref{Ham2})
 exactly to
 obtain the many-body energy spectrum resolved by the total momentum at $k_w=1.4/l_B$, 
 as shown in Fig.~\ref{Fig2}(a) and (c) for eight and twelve electrons, respectively.
 We see that the low energy spectrum is similar to that of half-filled lowest Landau level in two 
 dimensions, the paradigmatic CFL~\cite{dong2023PRL,goldman2023PRL} (see Appendix D),  
 suggesting that the phase we find in Weyl semimetals is also the CFL.
 Explicitly, in Fig.~\ref{Fig2}(a), the twelve states with lowest energy 
 (highlighted by the red color and each red point represents doubly degenerate states) 
 appear in the same momentum sectors as the 2D CFL, which can be explained by the
 picture of stacking composite fermions~\cite{geraedts2018PRL,dong2023PRL}.
 For $n_e=12$ ($n_e$ denotes the number of electrons) in Fig.~\ref{Fig2}(c), 
 we mark the states with lowest energy (each is two-fold degenerate) 
 in the momentum sectors where the ground states of the 2D CFL lie by the red color.
 We see that in two of them the states colored red are ground states of the system.
 In another one, 
 the energy of the states colored red is a little higher than a few states in other momentum sectors, 
 which is possible in a finite-size system considering the gapless nature of the CFL.

To provide further evidence showing that the system is in the CFL phase, 
we calculate the guiding center static structure factor~\cite{rezayi1999PRL,haldane2000PRL,jian2020PRR} 
(also see Appendix E)
\begin{equation}\label{SF}
	S(\bm{q})=\frac{\langle\overline{\rho}(\bm{q})\overline{\rho}(-\bm{q}) \rangle}{n_e}
	-\frac{\langle\overline{\rho}(0)\rangle^2}{n_e},
\end{equation} 
where $\overline{\rho}(\bm{q})=\sum_ie^{i\bm{q}\cdot\bm{R}_i}$ with $\bm{R}_i$ being the 
guiding center coordinate operator of the $i$th electron,
$\bm{q}=q_1\bm{g_1}+q_2\bm{g_2}$ with $q_1,q_2$ taking integer values, and $\langle...\rangle$
indicates the expectation value for the ground state. 
We plot the $S(\bm{q})$ in Fig.~\ref{Fig2}(b) and (d) for $n_e=8$ and 12, respectively, and
indicate the magnitude of the Fermi wave vector of free electrons $k_F$ from the origin by an inner dashed red line and $2k_F$
by an outer dashed line.
We see that $S(\bm{q})$ exhibits strong peaks on a closed loop near |$\bm{q}|=2k_F$, 
indicating the Fermi surface of 
composite fermions~\cite{geraedts2016Sciencea,jian2020PRR}.
Note that the composite fermion Fermi surface for $n_e=12$ is deformed from an exact circle,
which may be attributed to the anisotropy of the Weyl semimetal Hamiltonian (\ref{Ham1}).

 \section{Moore-Read states}
 We show the energy spectrum of the Hamiltonian (\ref{Ham2}) at 
 $k_w=2.1/l_B$ in Fig.~\ref{Fig3}(a) for ten electrons and (c) for twelve electrons.
 In both cases, we see six states with lowest energy (highlighted by red color and states in 
 Fig.~\ref{Fig3}(c) are two-fold degenerate) lying in the exact momentum sector predicted by the
 $(2,4)$ counting rule of MR states~\cite{bernevig2008PRL}, i.e. there are at most two electrons in four consecutive orbitals.
 To probe the excitation properties of the possible MR states,
 we calculate the particle entanglement spectrum~\cite{sterdyniak2011PRL,wang2025PRB}, i.e.  the eigenvalues of the reduced density matrix 
  \begin{equation}\label{ES}
  	\rho_A={\rm Tr}_B(\frac{1}{6}\sum_{i=1}^6 |\Psi_i\rangle \langle \Psi_i |) ,
  	\end{equation}
  	where the summation is over the six states with lowest energy.
The reduced density matrix is obtained by tracing out $n_b$ electrons in the $B$ part out of the total $n_e$ electrons with $n_a=n_e-n_b$ electrons left.
 The particle entanglement spectrum for $n_e=10$ and  $n_a=4$ is shown in Fig.~\ref{Fig3}(b). 
 We see a gap
 (indicated by the red dashed line) below which the numbers of eigenvalues in each momentum sector are 
 200, 196, 201, 196, 200, 196, 201, 196, $\dots$, consistent with the $(2,4)$ counting rule of the MR states.
 A clearer gap manifests in Fig.~\ref{Fig3}(d) for $n_e=12$ and  $n_a=4$ below which the numbers of eigenvalues 
 are 394 for momentum sectors with $\mod(k_1+N_1k_2,8)=0$, 396 for $\mod(k_1+N_1k_2,8)=2$ and 
 384 for others, also satisfying the counting rule of MR states.
 Thus, the particle entanglement spectrum provides further evidence that the states are indeed the MR states.
 Interestingly, for the half-filled first excited Landau level in 2D with Coulomb interactions,
 commonly 
 regarded as the MR states, a gap below which the number of states satisfies the MR counting rule exists
 when $n_e=10$ and  $n_a=4$ but is absent when $n_e=12$ and  $n_a=4$~\cite{reddy2024PRL}.

\begin{figure}[t]
	\includegraphics[width=3.4in]{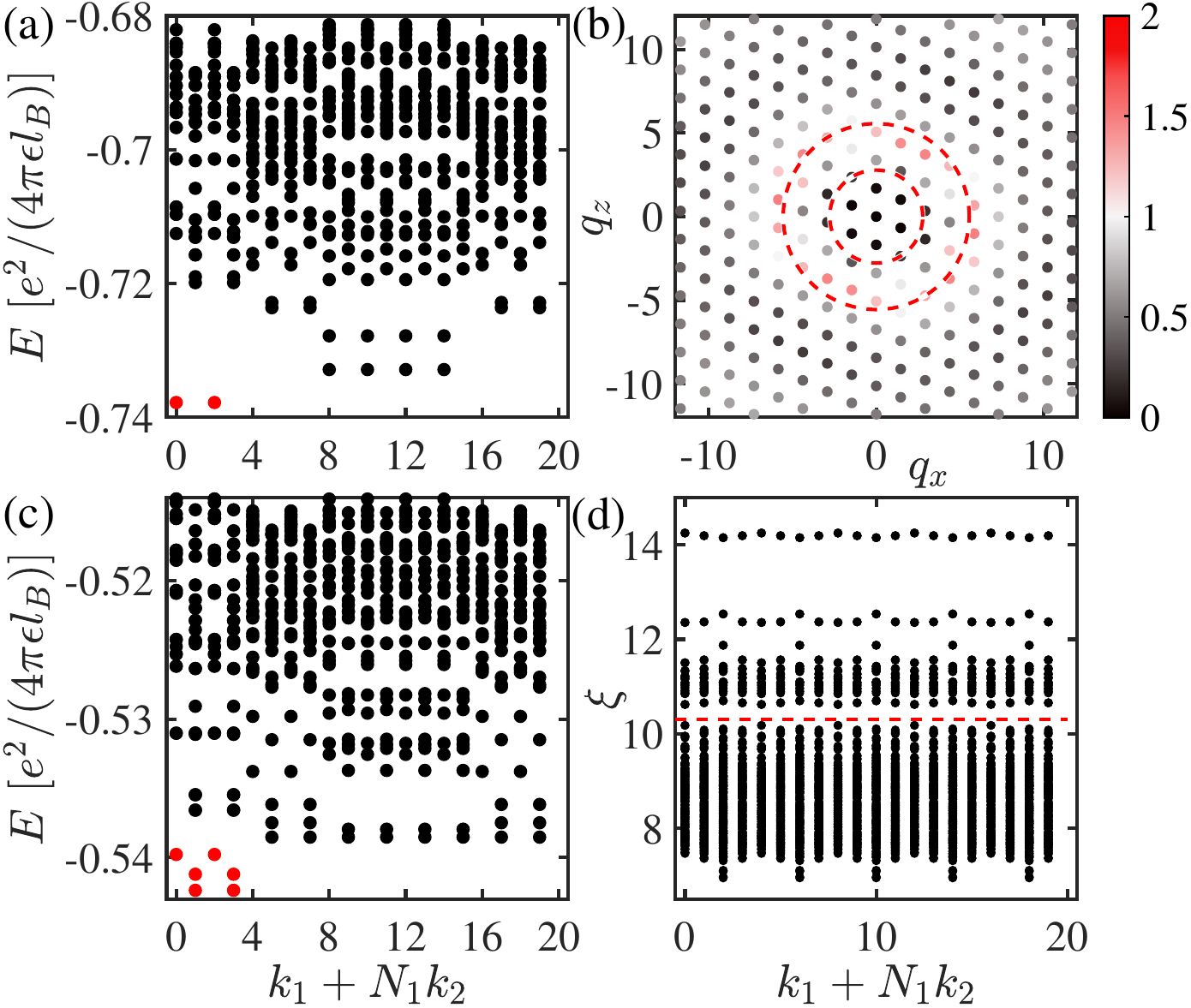}
	\caption{(a) The many-body energy spectrum and (b) the guiding center static structure factor 
		$S(\bm{q})$ of the ground state for ten electrons under vertical magnetic fields.
		In (a), the low energy states colored red are similar to these of the 2D CFL.
		(c) The many-body energy spectrum and (d) the particle entanglement spectrum for 
		ten electrons under a tilted magnetic field.	
		The number of states below the red dashed line agrees with the counting rule of the MR states.
		Here, we take $\widetilde{M}=M/l_B^2=0.05E_0,\widetilde{D}_1=D_1/l_B^2=0.025E_0,
		\widetilde{D}_2=D_2/l_B^2=0.005E_0,\widetilde{A}=A/l_B=0.05E_0$, and $k_w=1.7/l_B$.
	}	
	\label{Fig5}
\end{figure}
 
 \section{Charge density waves}
 We now provide evidence showing that the ground states 
 are CDWs when $k_w=2.7/l_B$. Figure~\ref{Fig4}(a) 
 displays the many-body energy spectrum  
 for eight electrons, illustrating quasi-degenerate ground states in momentum sectors with
 $k_1+N_1k_2=0$ and 2, each having four-fold degeneracy.  
 For twelve electrons as shown in Fig.~\ref{Fig4}(c),
 we see twelve quasi-degenerate ground states in $k_1+N_1k_2=2,10,18$ momentum sectors
  with four-fold degeneracy in each momentum sector.
 The high degeneracy seen here is often a signature of CDWs~\cite{rezayi1999PRL,rezayi2000PRL}.
 To elucidate the properties of the states, we calculate $S(\bm{q})$ and show 
 them in Fig.~\ref{Fig4}(b) and (d) for eight and twelve electrons, respectively.
 The figures exhibit two isolated sharp peaks, obviously evidencing a CDW phase.

 \section{Tilted magnetic field induced transitions}
 We now show that the transition from the CFL to the MR states can be induced by controlling the direction 
  of the magnetic field, which is more 
  experimentally accessible. 
  Under the vertical magnetic field along $y$, the two ground states [colored red in Fig.~\ref{Fig5}(a)] exist in the same 
  momentum sectors as the 2D CFL for ten electrons (see Appendix D).
  The structure factor $S(\bm{q})$ demonstrates
  a circle of peaks near |$\bm{q}|=2k_F$ [Fig.~\ref{Fig5}(b)], characteristic of the CFL.
  We then tilt the magnetic field such that the magnetic field attains a nonzero $x$ 
  component, i.e., $\bm{B}=(B_x, B_y, 0)$.
  When $B_x/B_y=1/25$, we see that the ground states become the MR states,
  confirmed by the many-body energy spectrum [Fig.~\ref{Fig5}(c)] and the particle entanglement spectrum 
  [Fig.~\ref{Fig5}(d)], where the number of
  states below the gap indicated by the red dashed line satisfies the MR counting rule. 
  The transition is also attributed to 
  the increase of the number of nodes in the
  single-particle wave function of the considered Landau level (see Appendix C).
 
   \begin{figure}[t]
 	\includegraphics[width=3.4in]{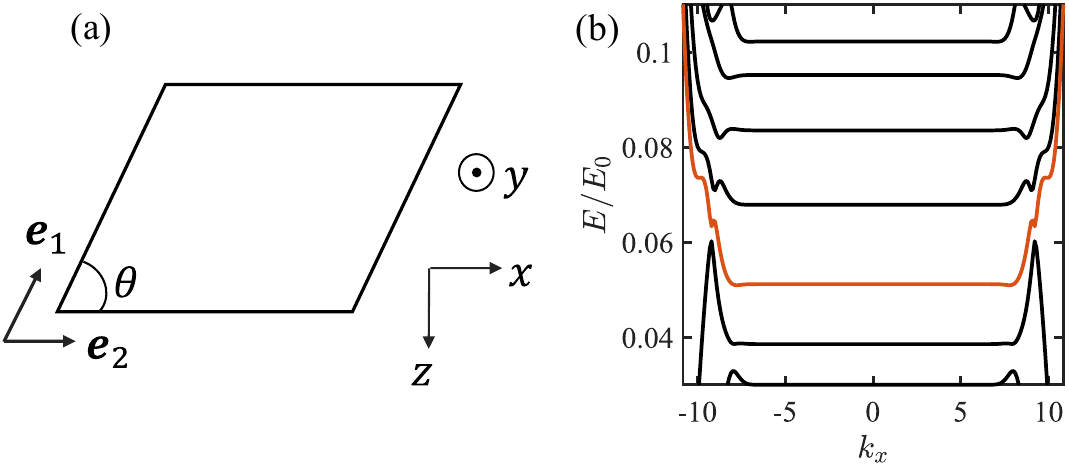}
 	\caption{
 		(a) The geometry of the sample viewed from directly above. The directions of boundaries are 
 		indicated by unit vectors $\bm{e}_1$ and $\bm{e}_2$ with an angle $\theta$.
 		(b) Energy spectra of the Landau levels arising from Weyl orbits when ${k_w=2.1/l_B}$ and other parameters are the same as Fig.~\ref{Fig1} in the main text.
 		We study the many-body physics on the Landau level marked by the red color, 
 		below which the total Chern number sums to zero.}
 	\label{FigS1}
 \end{figure}

 \section{Conclusion}
  In summary, we have demonstrated the existence of the CFL, MR and CDW phases 
  and their transitions in a Weyl-orbit-based Landau level.
  The transition is caused by a significant change in the wave function of the 
  Landau level states as the distance between two Weyl points varies.
  We also show that varying the direction of a magnetic field can induce the 
  transition from the CFL to MR states, providing an experimentally accessible method for
  observing the transition.
  In future, we will study the phase diagram under broader parameter ranges to investigate the possibility of other types of phase transitions, e.g., the $2k_F$ density wave instabilities due to gauge fluctuations~\cite{jian2020PRR}.
  The effect of Landau level mixing is also an interesting question, which is left for future work.

 \begin{acknowledgments}
 	We thank Y.-B. Yang and S.-C. Ma for helpful discussions.
 	This work is supported by the National Natural Science Foundation of China (Grant No. 12474265, 11974201)
 	and Innovation Program for Quantum Science and Technology (Grant No. 2021ZD0301604).
 	We also acknowledge the support by Center of High Performance Computing, Tsinghua University.
 \end{acknowledgments}

 \section{Appendix A: A Weyl semimetal under a magnetic field}\label{AppA}
 	\setcounter{equation}{0}
 \renewcommand{\theequation}{A\arabic{equation}}
 In this Appendix, we will review how Weyl orbits arise in a Weyl semimetal under a magnetic field~\cite{potter2014NC,wang2017PRL} 
 and show single-particle energy spectra of the Landau level based on Weyl orbits.

 We follow Fig.~\ref{Fig1}(a) in the main text to schematically demonstrate how an electron evolves in a Weyl semimetal under a magnetic field.
 Specifically, 
 starting from the projection of a Weyl point onto one surface, say the blue point on the top surface, 
 an electron travels along the Fermi arc subjected to the magnetic field.
 At the end of the Fermi arc, a red point, it tunnels to the bottom surface through the Weyl point and 
 then moves along the Fermi arc to the blue point.
 Subsequently, it tunnels back to the top surface, forming a closed loop called a Weyl orbit.
 
 The Weyl orbit gives rise to Landau levels, as shown in Fig.~\ref{FigS1}(b),
 where we take OBCs in the $z$ direction and PBCs in the $x$ direction so that $k_x$ is a good quantum number.
 In this paper, we study the many-body physics on the Landau level highlighted by the red color, 
 given that the total Chern number below it sums to zero.
 Note that for numerical calculations, we choose the Landau gauge $\bm{A}=(Bz,0,0)$ for the magnetic field 
 $\bm{B}=(0,B,0)$.
 Since we focus on the many-body bulk physics, we use the following functions as a basis by considering a system that is infinite along $z$, 
 \begin{equation}\label{S1Basis}
 \begin{split}
 	\phi_{m,h}(y,z)=&\frac{1}{\sqrt{\sqrt{\pi}2^m m!l_B}}\sqrt{\frac{2}{L_y}}
 	\exp{\left[-\frac{(z-z_0)^2}{2l_B^2}\right]}\\
 	&\mathcal{H}_m(\frac{z-z_0}{l_B})
 	\sin\left[\frac{h\pi}{L_y}\left(y+\frac{L_y}{2}\right) \right],
 \end{split}
 \end{equation} 
 where $m$ takes non-negative integer values, $h$ takes positive integer values, 
 $l_B=\sqrt{\hbar/(eB)}$ is the magnetic length,
 $L_y$ is the thickness of the sample in the $y$ direction, $z_0=-k_xl_B^2$ is the guiding center, and
 $\mathcal{H}_m$ is the Hermite polynomial.

  \begin{widetext}
 \section{Appendix B: Construction of the many-body Hamiltonian}\label{AppB}
 	\setcounter{equation}{0}
 \renewcommand{\theequation}{B\arabic{equation}}
 In this Appendix, we will show how to construct the interacting Hamiltonian (\ref{Ham2})
 in the main text and present the procedure to arrive at the 2D Brillouin zone.
 
 In this paper, we consider a parallelogram geometry in the horizontal plane described by 
 unit vectors $\bm{e}_1$ and $\bm{e}_2$ with an angle $\theta=\pi/3$ as shown in Fig.~\ref{FigS1}(a). 
 The system's lengths along $\bm{e}_1$ and $\bm{e}_2$ are $L_1$ and $L_2$, respectively.
 In the study of the many-body physics of the half-filled Landau level in the main text,
 we take OBCs along $y$ and PBCs along $\bm{e}_1$ and $\bm{e}_2$.
 The single-particle state on the Landau level we investigate is a spinor 
 $\Psi_j=(\psi_{j,0},\psi_{j,1})^T$ where
 \begin{equation}\label{S2State}
 	\psi_{j,\sigma}=\sum_{{m,h}}c_{m,h,\sigma}\tilde{\phi}_{m,h,j}
 \end{equation} 
 with $\sigma=0,1$ and $c_{m,h,\sigma}$ obtained by numerically diagonalizing the single-particle
 Weyl semimetal Hamiltonian in the basis (\ref{S1Basis}).
 Here  
 \begin{equation}\label{S2Basis}
 	\begin{aligned}
 		\tilde{\phi}_{m,h,j}(\bm{r})&=\frac{1}{\sqrt{\sqrt{\pi}2^m m!l_BL_2}}\sqrt{\frac{2}{L_y}}
 		\sum_{n=-\infty}^{+\infty}\exp\left[i\frac{2\pi}{L_{2}}\big(j+nN_{\phi}\big)\big(x-\frac{\pi\cos\theta}{L_{2}\sin\theta}(j+nN_{\phi})l_{B}^{2}\big)\right]\\
 		&	\exp{\left[-\frac{(z-z_n)^2}{2l_B^2}\right]}\mathcal{H}_m(\frac{z-z_n}{l_B})
 		\sin\left[\frac{h\pi}{L_y}\left(y+\frac{L_y}{2}\right) \right]
 	\end{aligned}
 \end{equation} 
 is a remedied form of (\ref{S1Basis}) for consistency with PBCs,
 where $z_n=-2\pi jl_B^2/L_2-nL_1\sin \theta $, $j=0,...,N_\phi-1$ with $N_\phi$ being the degeneracy of the 
 Landau level.
 To be explicit, it is easy to obtain
 \begin{equation}\label{S2L1}
 	\tilde{\phi}_{m,h,j}(\bm{r}+L_1\bm{e}_1)=e^{i2\pi N_\phi x/L_2}e^{i2\pi^2\cos \theta N_\phi^2l_B^2/(L_2^2\sin \theta) }\tilde{\phi}_{m,h,j}(\bm{r}),
 \end{equation}
 \begin{equation}\label{S2L2}
 	\tilde{\phi}_{m,h,j}(\bm{r}+L_2\bm{e}_2)=\tilde{\phi}_{m,h,j}(\bm{r}).
 \end{equation}
 The magnetic translation operator is $\mathcal{T}(\bm{d})=e^{i\bm{d}\cdot\bm{K}}$ where
 $\bm{K}=-i\nabla+e\bm{A}-e\bm{B}\times \bm{r}$ for $\bm{d}=(d_x,0,d_z)$.
 We can also derive 
 \begin{equation}\label{S2MTO1}
 	\mathcal{T}(L_1\bm{e}_1)=e^{-i2\pi N_\phi x/L_2}e^{-i2\pi^2\cos \theta N_\phi^2l_B^2/(L_2^2\sin \theta) }T(L_1\bm{e}_1),
 \end{equation}
 and
 \begin{equation}\label{S2MTO2}
 	\mathcal{T}(L_2\bm{e}_2)=T(L_2\bm{e}_2)
 \end{equation}
 where $T(\bm{d})$ is the ordinary translation operator.
 As a result, the wave functions in Eq. (\ref{S2Basis}) satisfy
 \begin{eqnarray}
 	\mathcal{T}(L_1\bm{e}_1) \tilde{\phi}_{m,h,j}({\bm r}) &=& \tilde{\phi}_{m,h,j}({\bm r}) \\
 	\mathcal{T}(L_2\bm{e}_2) \tilde{\phi}_{m,h,j}({\bm r}) &=& \tilde{\phi}_{m,h,j}({\bm r}),
 \end{eqnarray} 
 which is consistent with the PBCs.
 In the derivation of (\ref{S2MTO1}) and (\ref{S2MTO2}), we have used the Zassenhaus formula that 
 $e^{A+B}=e^Ae^Be^{-[A,B]/2}$ 
 when $[[A,B],A]=[[A,B],B]=0$ with $[A,B]$ being the commutator of $A$ and $B$.
 
 By projecting the Coulomb interaction onto one Landau level, 
 we obtain the interacting Hamiltonian
 \begin{equation}\label{S2Ham1}
 	\hat{H}_I=\sum_{j_1,j_2,j_3,j_4=0}^{N_\phi-1}B_{j_1j_2j_3j_4}\hat{a}^\dagger_{j_1}\hat{a}^\dagger_{j_2}
 	\hat{a}_{j_3}\hat{a}_{j_4},	
 \end{equation}
 where $\hat{a}^\dagger_{j_\mu}$ ($\hat{a}_{j_\mu}$) is the creation (annihilation) operator of the
 single-particle state $\Psi_j$ in the Landau level under investigation.
 The coefficients
 \begin{equation}\label{S2B}
 	B_{j_1j_2j_3j_4}=\sum_{\sigma_1\sigma_2}\sum_{m_{1,2,3,4}h_{1,2,3,4}}c_{m_1,h_1,\sigma_1}^{*}
 	c_{m_2,h_2,\sigma_2}^{*}c_{m_3,h_3,\sigma_2}c_{m_4,h_4,\sigma_1}
 	A_{j_1j_2j_3j_4}^{m_1h_1m_2h_2m_3h_3m_4h_4},
 \end{equation}
 where 
 \begin{eqnarray}\label{S2}
 	A_{j_1j_2j_3j_4}^{m_1h_1m_2h_2m_3h_3m_4h_4}=\frac{e^2}{4\pi \epsilon}\frac{1}{2L_xL_z\sin \theta}
 	\sum_{q_x,q_z,s,t}^\prime \frac{2\pi}{q}F({\bm q})G_1({\bm q})G_2(\bm q)
 	e^{i\frac{2\pi s(j_3-j_1)}{N_\phi}} \nonumber \\ 
 	\times \delta_{q_x,\frac{2\pi}{L_x}t}
 	\delta_{q_z,\frac{2\pi}{L_z\sin\theta}s-\frac{2\pi\cos \theta}{L_x\sin\theta}t}
 	\delta_{t,j_1-j_4}^\prime
 	\delta_{j_1+j_2,j_3+j_4}^\prime,
 \end{eqnarray}	
 with 
 \begin{equation}\label{S2F}
 	F(\bm{q})=\int_{-L_y/2}^{L_y/2} \int_{-L_y/2}^{L_y/2}{\rm d}y_{1}{\rm d}y_{2}(\frac{2}{L_{y}})^{2}\sin[\frac{h_{1}\pi}{L_{y}}(y_{1}+\frac{L_{y}}{2})]\sin[\frac{h_{4}\pi}{L_{y}}(y_{1}+\frac{L_{y}}{2})]\sin[\frac{h_{2}\pi}{L_{y}}(y_{2}+\frac{L_{y}}{2})]\sin[\frac{h_{3}\pi}{L_{y}}(y_{2}+\frac{L_{y}}{2})]e^{-q|y_{1}-y_{2}|},
 \end{equation}
 \begin{equation}\label{S2G1}
 	G_1({\bm q})=\sqrt{\frac{\min{(m_1,m_4)}!}{\max{(m_1,m_4)}!}} e^{-q^2l_B^2/4}
 	[\frac{\mbox{sgn}(m_{1}-m_{4})q_{x}+iq_{z}}{\sqrt{2}}l_B]^{|m_{1}-m_{4}|}L_{\min(m_{1},m_{4})}^{(|m_{1}-m_{4}|)}(\frac{q^{2}l_B^{2}}{2}),		
 \end{equation}
 \begin{equation}\label{S2G2}
 	G_2({\bm q})=\sqrt{\frac{\min{(m_2,m_3)}!}{\max{(m_2,m_3)}!}} e^{-q^2l_B^2/4}
 	[\frac{-\mbox{sgn}(m_{2}-m_{3})q_{x}-iq_{z}}{\sqrt{2}}l_B]^{|m_{2}-m_{3}|}L_{\min(m_{2},m_{3})}^{(|m_{2}-m_{3}|)}(\frac{q^{2}l_B^{2}}{2}).		
 \end{equation} 
 Here $\bm{q}=(q_x,q_z)$, $q=|\bm{q}|$, $L_{\min(m_{1},m_{4})}^{(|m_{1}-m_{4}|)}$ is the Laguerre polynomial and $s,t$ take integer values.
 The detailed derivation can be found in the Appendix of  Ref.~\cite{wang2025PRB}.
 
 In the following, we will follow Ref.~\cite{wu2013PRL} to construct a 2D Brillouin
 zone for Landau 
 level states, i.e., rewrite the Hamiltonian (\ref{S2Ham1}) as
 \begin{equation}\label{S2Ham2}
 	{\hat{H}_I=\sum_{\bm{k}_1,\bm{k}_2,\bm{k}_3,\bm{k_4}}C_{\bm{k}_1\bm{k}_2\bm{k}_3\bm{k}_4}\hat{c}^\dagger_{\bm{k}_1}\hat{c}^\dagger_{\bm{k}_2}
 		\hat{c}_{\bm{k}_3}\hat{c}_{\bm{k}_4},}
 \end{equation} 
 where $\hat{c}_{\bm{k}}$ ($\hat{c}^\dagger_{\bm{k}}$) is the annihilation (creation) operator of
 the single-particle state with two labels $k_1,k_2$ for each $\bm{k}$ with $k_i=0,1,...,N_i-1$ for $i=1,2$ and $N_1\cdot N_2=N_\phi$.
 It is nothing more than a basis transformation.
 We define
 \begin{equation}\label{S2BT}
 	\hat{c}^\dagger_{\bm{k}}=\frac{1}{\sqrt{N_{1}}}\sum_{n=0}^{N_{1}-1}e^{i2\pi nk_{1}/N_{1}}\hat{a}^\dagger_{j=nN_{2}+k_{2}},
 \end{equation}
 and utilizing the explicit form of $\tilde{\phi}_{m,h,j}(\bm{r})$ in (\ref{S2Basis}), we find
 \begin{equation}\label{S2Bloch1}
 	\mathcal{T}(L_{1}\bm{e}_1/N_{1})	\hat{c}^\dagger_{\bm{k}}|0\rangle=e^{i2\pi k_{1}/N_{1}}\hat{c}^\dagger_{\bm{k}}|0\rangle,
 \end{equation}
 \begin{equation}\label{S2Bloch2}
 	\mathcal{T}(L_{2}\bm{e}_2/N_{2})	\hat{c}^\dagger_{\bm{k}}|0\rangle=e^{i2\pi k_{2}/N_{2}}\hat{c}^\dagger_{\bm{k}}|0\rangle,
 \end{equation}
 which is like the Bloch states of a 2D lattice system with $N_1\times N_2$ unit cells
 and lattice constant $L_1/N_1,L_2/N_2$ in the $\bm{e}_1$ and $\bm{e}_2$ directions, respectively. 
 We thus can construct 
 a 2D Brillouin zone with Bloch wave vectors $\bm{k}=k_1\bm{g}_1+k_2\bm{g}_2$,
 where $\bm{g}_i\cdot \bm{e}_j=\delta_{ij}2\pi/L_i $
 with $\delta_{ij}$  being the Kronecker delta function.
 The reciprocal lattice vector $\bm{G}_1=N_1\bm{g}_1,\bm{G}_2=N_2\bm{g}_2$, 
 so $|\bm{G}_1|=|\bm{G}_2|$ when $L_1/N_1=L_2/N_2$, corresponding to a hexagonal Brillouin zone when
 $\theta=\pi/3$.
 We can easily find that
 \begin{equation}\label{S2C}
 	C_{\boldsymbol{k}_{1}\boldsymbol{k}_{2}\boldsymbol{k}_{3}\boldsymbol{k}_{4}}=\frac{1}{N_{1}^{2}}\sum_{n_{1},n_{2},n_{3},n_{4}=0}^{N_{1}-1}e^{-i2\pi(n_{1}k_{1,1}+n_{2}k_{2,1}-n_{3}k_{3,1}-n_{4}k_{4,1})/N_{1}}A_{n_{1}N_{2}+k_{1,2},n_{2}N_{2}+k_{2,2},n_{3}N_{2}+k_{3,2},n_{4}N_{2}+k_{4,2}},
 \end{equation}
 where $k_{i,j}$ stands for the $k_j$ component of $\bm{k}_i$.
 \end{widetext}
 
 \section{Appendix C: Real-space wave functions in a Landau level arising from Weyl orbits  }\label{AppC}
 	\setcounter{equation}{0}
 \renewcommand{\theequation}{C\arabic{equation}}
 In this Appendix, we will display the real-space wave functions in a Weyl-orbit-based Landau
 level that we focus on to elucidate the mechanism of the transitions between the many-body phases in the 
 main text. 
 
  \begin{figure*}[h]
 	\includegraphics[width=6.5in]{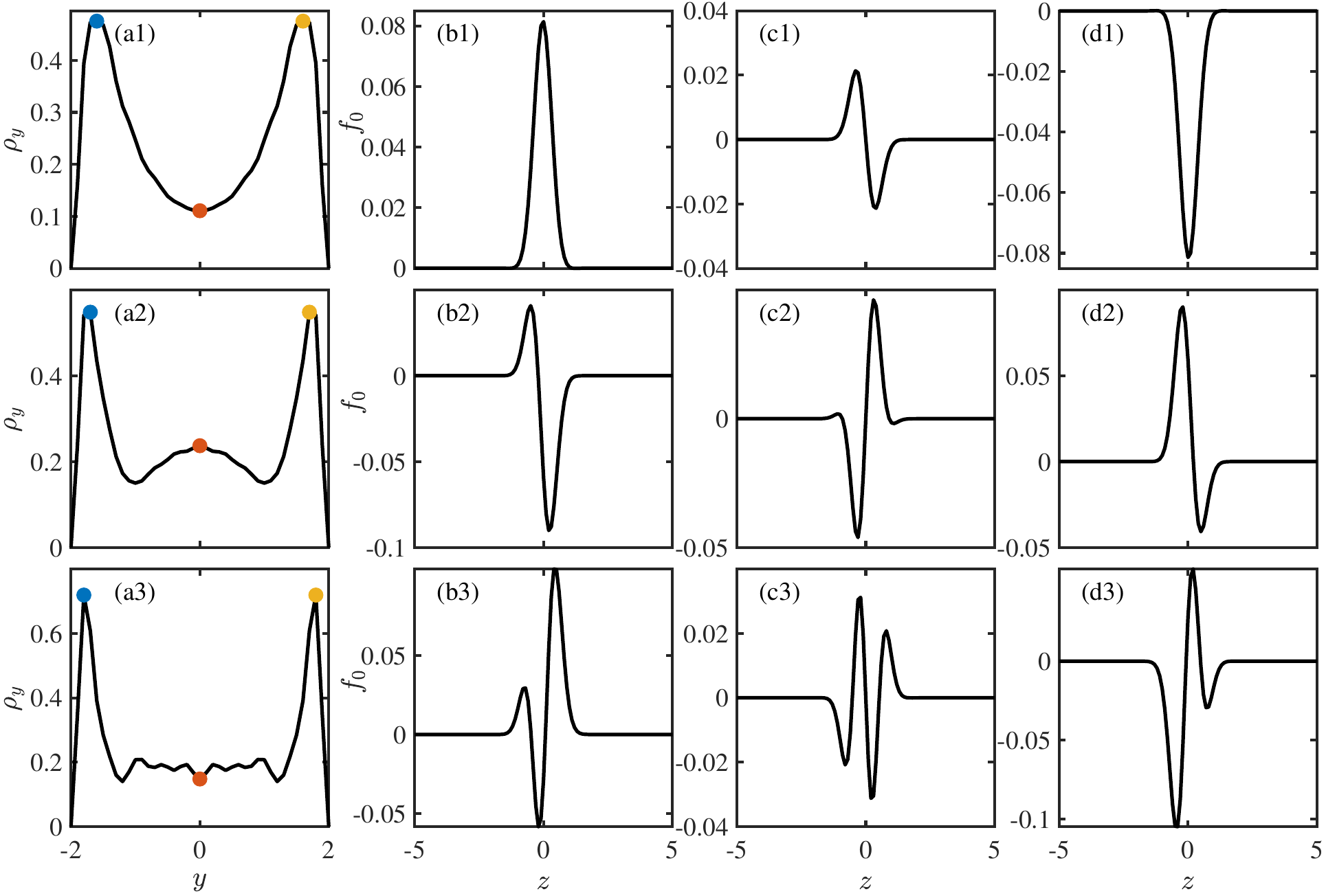}
 	\caption{
 		(a1)-(a3) The electron density distribution $\rho_y$ with respect to $y$.
 		The real-space wave function $f_0(y_0,z)$ versus $z$ (b1)-(b3) near 
 		a bottom surface, (c1)-(c3) deep in the bulk, (d1)-(d3) near a top surface,
 		at a fixed $y_0$ indicated by the blue, red, and yellow points in (a), respectively.
 		In the first, second, and third rows, $k_w=1.4/l_B$, $k_w=2.1/l_B$, and $k_w=2.7/l_B$, respectively. 
 		Other parameters are the same as Fig.~\ref{Fig1} in the main text.
 	}
 	\label{FigS4}
 \end{figure*}
 \begin{figure*}[h]
 	\includegraphics[width=6.5in]{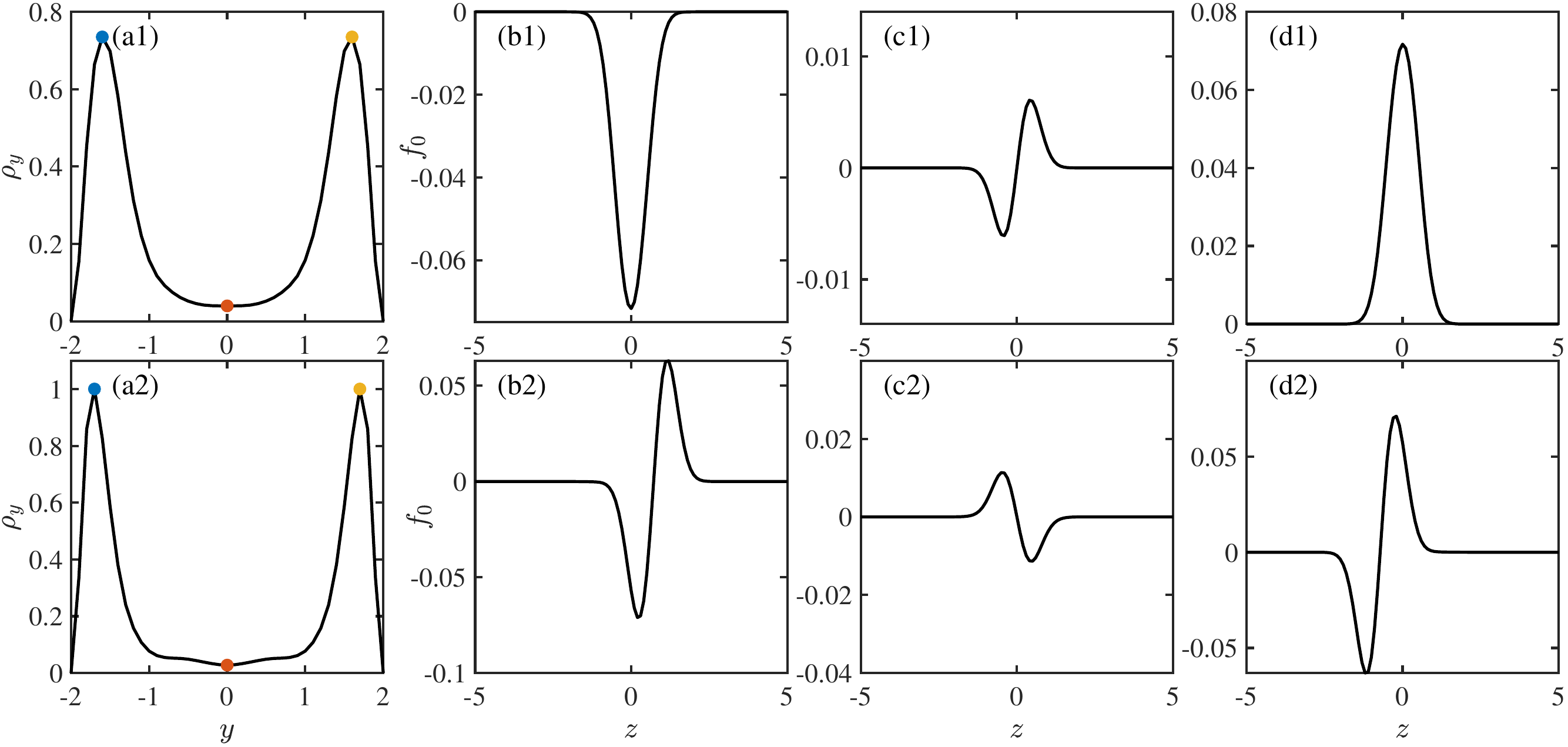}
 	\caption{
 		The same as Fig.~\ref{FigS4} except that we take the parameters the same as Fig.~\ref{Fig5} in the main 
 		text with a vertical magnetic field in the first row and a tilted magnetic field in the second row.
 	}
 	\label{FigS5}
 \end{figure*}
 
 We first show that wave functions in the Landau level are mainly localized near top and bottom 
 surfaces. Without loss of generality, we consider $k_x=0$ such that the guiding center is located 
 at $z_0=0$. The density of the single-particle wave function $\bm{f}=(f_0,f_1)$ along $y$ is
 given by
 \begin{equation}\label{S5Density}
 	\rho_y=\int{\rm d}z(|f_0(y,z)|^2+|f_1(y,z)|^2),
 \end{equation} 
 where the wave function is described by
 \begin{equation}\label{S5State}
 	f_{\sigma}(y,z)=\sum_{m,h}c_{m,h,\sigma}\phi_{m,h}(y,z),
 \end{equation} 
 with $\phi_{m,h}$ being the wave function in Eq.~(\ref{S1Basis}).
 The density profiles shown in Figs.~\ref{FigS4}(a1)-\ref{FigS4}(a3) imply that 
 the states are mainly distributed near top and bottom surfaces.
 
 To demonstrate why the phase transition occurs as the distance between Weyl points is varied, 
 we plot $f_0(y,z)$ (similarly for $f_1$) with respect to $z$ at a fixed $y$ (near top and bottom surfaces) in 
 Fig.~\ref{FigS4} for different $k_w$'s.
 
 Figures~\ref{FigS4}(b1) and (d1) illustrate that at $k_w=1.4/l_B$, the wave functions near the surfaces,
 where the electron density is concentrated [see Fig.~\ref{FigS4}(a)], exhibit no nodes,
 similar to the lowest Landau level in 2D.
 Thus, the CFL is favored.
 Note that although the wave function in the bulk hosts one node [see Fig.~\ref{FigS4}(c1)], 
 it plays a much less important role because the electron density deep in the bulk is small 
 [see Fig.~\ref{FigS4}(a)].
 
 As we increase $k_w$ to $2.1/l_B$, we see from Figs.~\ref{FigS4}(b2) and \ref{FigS4}(d2) 
 that the wave function exhibits one node akin to the first excited Landau level in 2D, 
 suggesting that the MR states are favorable.
 
 At $k_w=2.7/l_B$, two nodes show up in the wave function on the
 surfaces [see Figs.~\ref{FigS4}(b3) and \ref{FigS4}(d3)], 
 supporting the CDW~\cite{rezayi1999PRL,rezayi2000PRL,haldane2000PRL}.
 
 Similarly, when we tune the magnetic field from a vertical one to a tilted one, we find that
 the wave functions on the surfaces develop a node as shown in Fig.~\ref{FigS5},
 leading to the transition from the CFL to MR states.

 \section{Appendix D: Many-body energy spectrum of the CFL and MR states in 2D}\label{AppD}
\begin{figure*}
 	\includegraphics[width=6.5in]{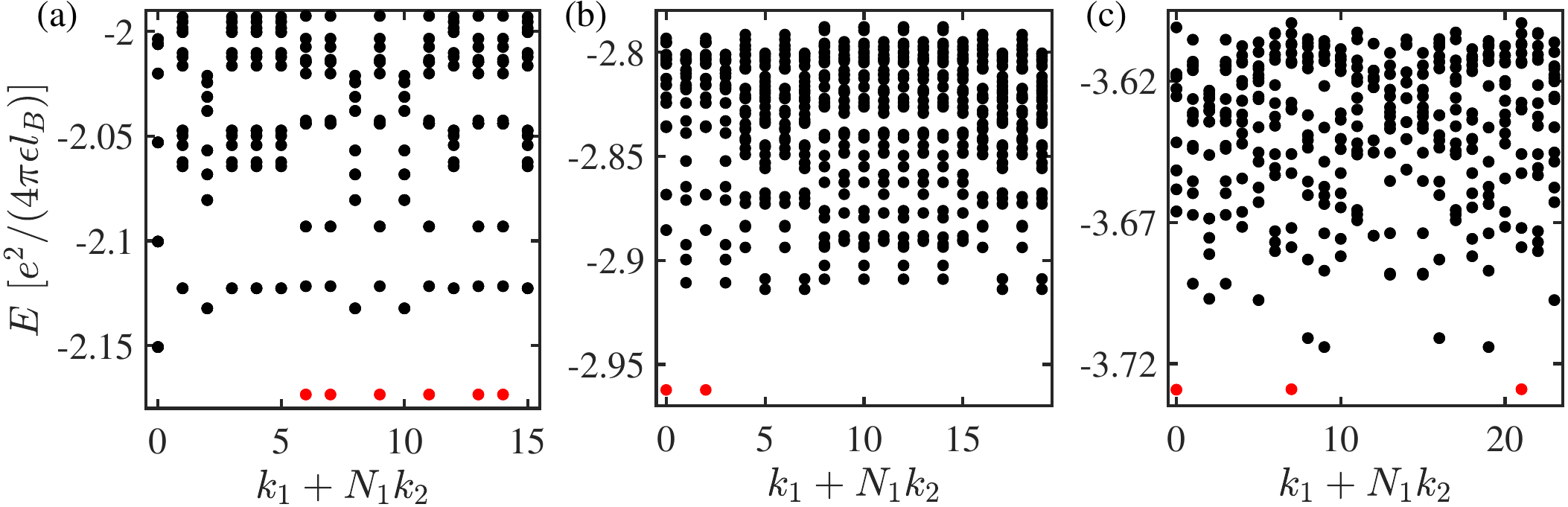}
 	\caption{The many-body energy spectra of the half-filled lowest Landau level in a 2D electron gas
 		with Coulomb interactions for (a) eight, (b) ten, and (c) twelve electrons, 
 		which indicate the paradigmatic CFL phase.
 		The lowest energy states are colored red. 
 	}
 	\label{FigS4_1}
 \end{figure*}

\begin{figure*}
 	\includegraphics[width=4.5in]{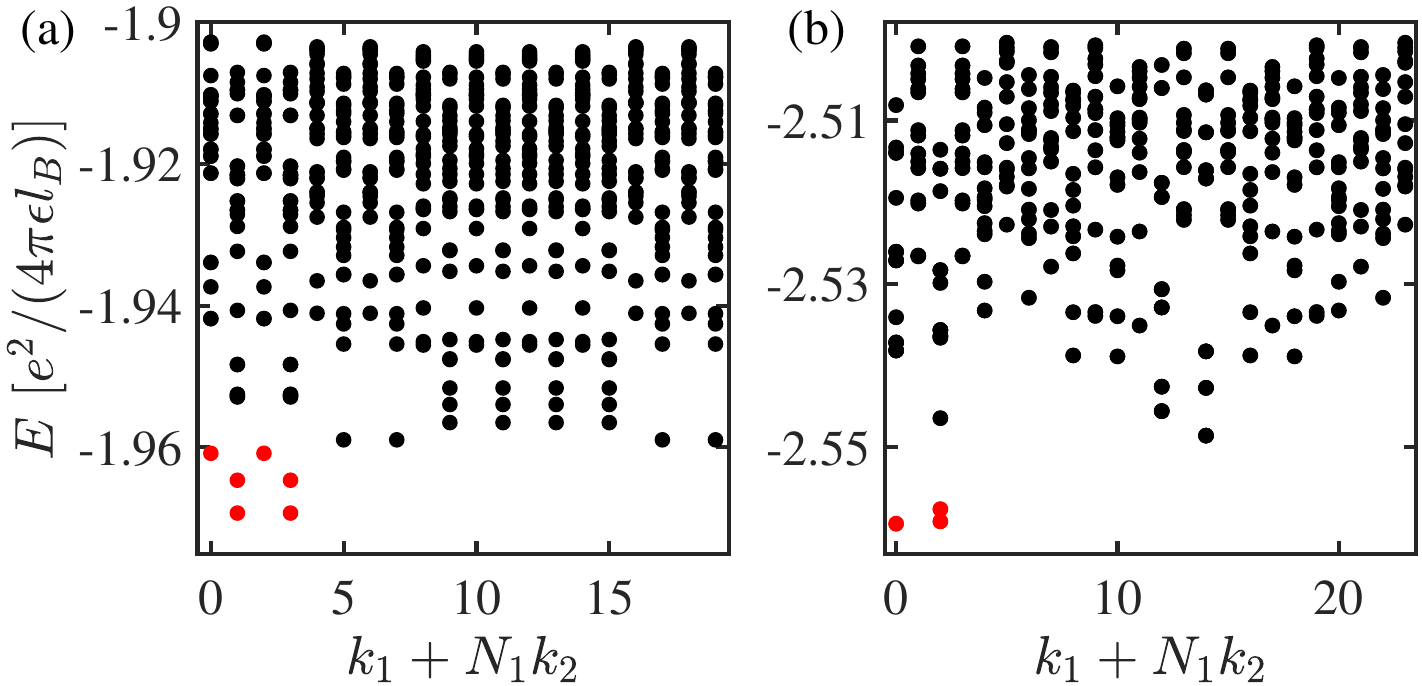}
 	\caption{The many-body energy spectra of the half-filled first excited Landau level in a 2D electron gas
 		with Coulomb interactions for (a) ten and (b) twelve electrons, 
 		hosting the MR states, which are colored red.
 	}
 	\label{FigS4_2}
 \end{figure*}

 	\setcounter{equation}{0}
 \renewcommand{\theequation}{D\arabic{equation}}
 In this Appendix, we show the many-body energy spectrum of the half-filled lowest and first 
 excited Landau level in a 2D electron gas with Coulomb interactions, 
 which host CFL (Fig.~\ref{FigS4_1}) and MR phases (Fig.~\ref{FigS4_2}), respectively.
 Here we project the Coulomb interaction onto a single Landau level. In the main text, 
 we have compared the many-body energy spectra in Weyl semimetals with those 
 in 2D electron gases, suggesting that the found phases are CFL or MR states. 

 
 \section{Appendix E: Guiding center coordinate operator}\label{AppE}
 	\setcounter{equation}{0}
 \renewcommand{\theequation}{E\arabic{equation}}
 In this Appendix, we will follow Ref.~\cite{jain2007} to provide a brief review on the guiding center coordinate operator.
 
 For simplicity, we consider a 2D electron gas under a magnetic field with the 
 Hamiltonian
 \begin{equation}\label{S3Ham1}
 	\hat{H}=\frac{1}{2m}(\hat{\pi}_x^2+\hat{\pi}_y^2),
 \end{equation} 
 where $\hat{\pi}_a=\hat{p}_a+e\hat{A}_a$ with $a=x,y$ and $\hat{\bm{A}}$ being the vector potential.
 We write the real space coordinate operator $\hat{r}^a$ as
 \begin{equation}\label{S3RC}
 	\hat{r}^a=\hat{R}^a+\hat{\tilde{R}}^a,
 \end{equation} 
 where $\hat{\tilde{R}}^a := (l_{B}^{2}/\hbar) \sum_{b}\epsilon^{ab}\hat{\pi}_{b}$, with
 $\epsilon^{ab}$ being the anti-symmetric Levi-Civita symbol.
 Physically, $\hat{\bm{R}}=(\hat{R}^x,\hat{R}^y)$ is the guiding center or origin of the cyclotron motion of the electron under
 the magnetic field, and $\hat{\tilde{\bm{R}}}=(\hat{\tilde{R}}^a,\hat{\tilde{R}}^b)$ is the coordinate relative to 
 the guiding center during the cyclotron motion.
 Based on the fact that
 $[\hat{\pi}_{x},\hat{\pi}_{y}]=-i\hbar^{2}/l_{B}^{2}$, we obtain the following commutation rules:
 \begin{equation}
 	\begin{aligned}
 		[\hat{R}^{a},\hat{\tilde{R}}^b]=&0, \\
 		[\hat{R}^{a},\hat{R}^b]=&\epsilon^{ab}i l_{B}^{2}, \\
 		[\hat{\tilde{R}}^a,\hat{\tilde{R}}^b]=&-\epsilon^{ab}il_{B}^{2}. 
 	\end{aligned}
 \end{equation}
 We see that $\hat{R}^{a}$ and $\hat{\tilde{R}}^a$ are two independent sets of canonical conjugate variables so 
 that we can define two sets of ladder operators:
 \begin{eqnarray}\label{S3Crea1}
 	\hat{b}^\dagger=\frac{1}{\sqrt{2}l_B}(\hat{R}^{x}-i\hat{R}^{y}),	\\
 	\hat{b}=\frac{1}{\sqrt{2}l_B}(\hat{R}^{x}+i\hat{R}^{y}),
 \end{eqnarray} 
 and
 \begin{eqnarray}\label{S3Crea2}
 	\hat{f}^\dagger=\frac{1}{\sqrt{2}l_B}(\hat{\tilde{R}}^{x}+i\hat{\tilde{R}}^{y}), \\
 	\hat{f}=\frac{1}{\sqrt{2}l_B}(\hat{\tilde{R}}^{x}-i\hat{\tilde{R}}^{y}).
 \end{eqnarray} 
 The Hamiltonian (\ref{S3Ham1}) can thus be written in term of $\hat{f}$ and $\hat{f}^\dagger$ as
 \begin{equation}\label{S3Ham2}
 	\hat{H}=\hbar\omega(\hat{f}^\dagger \hat{f}+\frac{1}{2})
 \end{equation} 
 with $\omega=eB/m$.
 The $n$th eigenstate in the $m$th Landau level corresponding to the energy $E_m=\hbar\omega(m+1/2)$ is given by 
 \begin{equation}\label{S3State}
 	|m,n\rangle=\frac{(	\hat{f}^\dagger)^m(\hat{b}^\dagger)^n}{\sqrt{n!m!}}|0\rangle,
 \end{equation} 
 which shows that $\hat{\tilde{R}}^{a}$ controls the transition between different Landau levels
 and  $\hat{R}^{a}$ governs the dynamics within a single Landau level.
 This indicates that only the guiding center operator $\hat{R}^{a}$ comes into effect after the projection 
 onto a specific Landau level.

 In the main text, we calculate the guiding center static structure factor
 \begin{equation}\label{S3SF}
 	S(\bm{q})=\frac{\langle\overline{\rho}(\bm{q})\overline{\rho}(-\bm{q}) \rangle}{n_e}
 	-\frac{\langle\overline{\rho}(0)\rangle^2}{n_e}
 \end{equation}  
 using the guiding center density operator $\overline{\rho}(\bm{q})=\sum_ie^{i\bm{q}\cdot\hat{\bm{R}}_i}$.
 Under the single Landau level projection, $\rho({\bm{q}})=L_{n}(q^{2}l_{B}^{2}/2)e^{-q^{2}l_{B}^{2}/4}\overline{\rho}({\bm{q}})$~\cite{park2014PRB},
 where $L_{n}$ is the Laguerre polynomial, and $\rho(\bm{q})=\sum_ie^{i\bm{q}\cdot\hat{\bm{r}}_i}$ 
 is the ordinary density operator, which is useful for the derivation of the second-quantization form of  $\rho({\bm{q}})$ for numerical 
 evaluations of $S(\bm{q})$.

  \bibliography{HalfFilled.bib}


\end{document}